# Extragalactic Planetary Nebulae (xPNe) – Chemical evolution and assembly histories of nearby galaxies using Oxygen and Argon abundances. From the local universe to cosmic dawn

Magda Arnaboldi[1], Ortwin Gerhard[2], Martin Roth[3], Souradeep Bhattacharya[4], Johanna Hartke[5], Chiara Spiniello[1], Azlizan Soemitro[3], Claudia Pulsoni[2], Lucas Valenzuela[6]

[1] ESO, Garching Germany
[2] MPE, Garching, Germany
[3] AIP, Potsdam, Germany
[4] CAR, Univ. Hertfordshire, UK
[5] FINCA, Turku, Finland
[6] LMU, Germany

**How galaxies formed and evolved in the expanding Universe is the main science goal of Near-Field Cosmology research. Studies of the properties of galaxies' resolved stars open a widow on their ancient galactic components, probing star-formation during epochs more than 10 billion years ago. Extragalactic Planetary Nebulae (xPNe) can help decipher the signatures of mergers and interactions persisting over many dynamical times by tracing elemental abundances coupled with their kinematics and spatial distributions. With new facilities, reaching higher angular resolution, area coverage and sensitivity, one can use xPNe to map Oxygen and Argon element abundances, in addition to their kinematics, to extend the Galactic Archeology investigation to the oldest stellar aggregates in our Local Universe.**

## Introduction

One can gain information on the motions of the stars in the low surface brightness regions of any galaxy by studying xPNe which are discrete tracers of their parent stellar population in the halo. xPNe are the glowing expanding shells (Arnaboldi et al. 2008) of gas and dust observed around stars that have recently left the asymptotic giant branch (AGB) and are evolving towards the white dwarf stage. The duration of the PN phase is in the range $1 \times 10^3$ to $3 \times 10^4$ yrs. They are traditionally considered the late phases of stars with masses between 0.7 and 8 $M_\odot$ but have been shown to exhibit a wide variety of striking morphologies pointing towards a binary evolution in many systems (Jones & Boffin 2017). Since the timescales between the AGB and PN phases are short, the spatial distribution and kinematics of PNe are expected to be identical to their parent population, having the same rotation and angular momentum as those of the stars (e.g. Arnaboldi et al. 1996,1998; Pulsoni et al. 2023). Studying xPNe as a population provides insight into galaxy structure and evolution.

Because of their relatively strong [O III] 5007 Å emission, PNe can be readily identified and are useful tracers for constraining the kinematics (Aniyan et al. 2018, 2021) and chemical abundances (Magrini et al. 2016; Stanghellini & Haywood 2018) over a large radial range, also in nearby galaxies of different morphological types (e.g., Cortesi et al. 2013; Pulsoni et al. 2018; Hartke et al. 2022). Elemental abundances of PNe shed light on the interstellar medium (ISM) conditions at the time of formation of their parent stellar population (Arnaboldi et al. 2022). When the PN ages are also constrained, it becomes possible to map abundance variations across different epochs of star formation in galaxies. Abundance distributions and gradients in galaxies were measured using PNe (Magrini et al. 2016; Kwitter & Henry 2022). Figure 1 below shows the spectrum of an xPN in the Andromeda M31 galaxy, with the emission lines which allow direct temperature determination and measurements of the element abundances for Oxygen and Argon, among other elements.

## Extragalactic PNe as tracers of chemical evolution

The formation history of any galaxy leaves imprints on the chemical composition of its stars. Of particular interest for nearby and distant galaxies is the measurement of [α/Fe] versus [Fe/H]. Because Type Ia Supernovae (SNe) produce primarily iron peak elements, while the more massive core-collapse SNe characterize α-rich bursts of star formation, by pairing Fe with one or more α-elements (e.g. Mg, Si, O), the Tinsley-Wallerstein diagram is used to decipher the chemical evolution and star formation history of stellar systems, *by using [Fe/H] as a clock and [α/Fe] as a star formation rate tracer*. In a population with a steady or decreasing star formation rate, this diagram shows an initial (high) level of [α/Fe] decreasing as the SN Ia kick-in. Thus far [α/Fe] vs [Fe/H] diagrams were measured from the spectra of individual red giant branch (RGB) stars or integrated galaxy spectra[1].

In Arnaboldi et al. (2022), by measuring direct abundances of Oxygen vs Argon, via detection of the temperature sensitive [OIII] 4363Å line and other strong emission lines, they found that the [log(O/Ar)] abundance ratios versus the Argon [Ar/H] abundance plane for PNe and HII regions in M31 is analogous to the [α/Fe] vs [Fe/H] for stars (see Arnaboldi et al. 2022 and Kobayashi et al. 2023). In the region of the M31 disc at R<14 kpc, the older, low-extinction, PNe follow a linear chemical enrichment with decreasing O and increasing Ar, while the younger, high-extinction, PNe lie below such track, with a near-constant log(O/Ar) value. Such a distribution, at younger ages, is an indication of a secondary gas-infall event. Galactic chemical evolution (GCE) models with a secondary infall event ~3 Gyr ago (with primordial metallicity) reproduce the PN distribution in the O-Ar plane with a characteristic loop pattern, see Figure 2. The comparison of the GCE model with measured O-Ar abundance seems to indicate that the O-Ar plane is potentially more sensitive to probing the merger/accretion/pollution events of the ISM. Following this first ground breaking example using xPNe in M31, the log(O/Ar) ratio vs [Ar/H] abundance diagnostic tool has been successfully applied to constrain the chemical enrichment of the ionized ISM in star forming galaxies at high redshift, from z=2 and up to nearly 8 (see Bhattacharya et al. 2025a) and in an extended sample of star forming galaxies at low redshift (z ~ 0.3; see Bhattacharya et al. 2025b).

## Opportunities for Extragalactic PNe as chemical tracers with new facilities

With the next generation of ESO facilities, the use of the log(O/Ar) vs Argon abundance plane as diagnostic for the chemical enrichment of the nebular phases of the stellar evolution will extend the galactic chemical evolution models from ``standard" resolved stellar populations diagnostic, i.e. absorption line spectroscopy of individual RGBs, to emission line diagnostic. Both limits in galactocentric radius (typically to 1 $r_e$ for absorption line spectroscopy) and in distances (current PN chemistry is done mostly in Local Group galaxies) will be enlarged using xPNe and O-Ar plane diagnostic. By reaching the stellar populations with the oldest stars, e.g. elliptical galaxies, and those at low surface brightnesses in any galaxy, where the fossil records of past accretion/merging events last longer, we will uncover the mass assembly history and star formation from early epochs.

---

[1] In the latter case stellar population parameters are derived by fitting stellar population models.

A new ground-based facility combining a large collecting area (10+meter-class telescope) with a wide field (O(1 deg)), highly multiplexed multi-object spectrographs with moderate spectral resolution (R ~ 2000 – 5000) can deliver a SNR≥3-5 for faint fluxes (O($F_{4363}$) ~ 1 x $10^{-18}$ erg $cm^{-2}$ $s^{-1}$). Reaching these fluxes allows the detection of temperature sensitive auroral lines (e.g. [OIII] 4363 Å) for direct O-Ar abundance plane determinations using xPNe, in galaxies and their haloes, out to 10 Mpc.

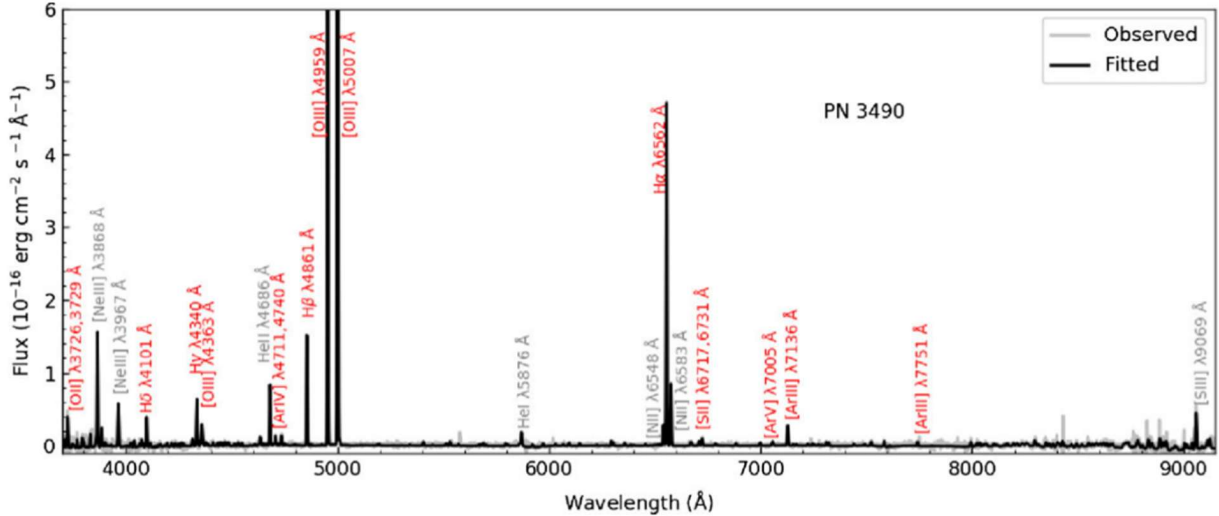

Figure 1: An example of the spectra observed by Hectospec for the xPN in M31. The spectra shown in grey is obtained following heliocentric correction, removal of sky-lines and flux calibration. The fitted spectra from Bhattacharya et al. (2022) is shown in black. The fluxes of emission lines both with red and grey label are tabulated in Table B1 in Bhattacharya et al. (2022). Same quality data can be acquired for xPNe in galaxies at 10+ Mpc

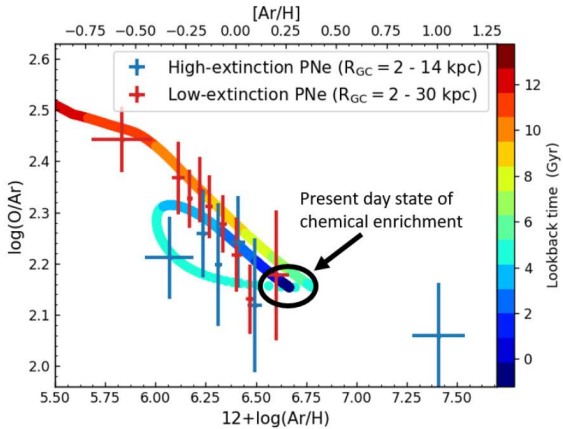

Figure 2: log(O/Ar) vs 12+ log(Ar/H) vs plane for xPNe in M31, binned by their 12+log(Ar/H) values. The high-extinction PNe (<2.5 Gyr old) are marked in blue while the low-extinction PNe (>4.5 Gyr old) are marked in red. The Galactic Chemical Evolution model, following chemical enrichment relations in Kobayashi et al. (2020), is colored by lookback time. Adapted from Arnaboldi et al. (2022).